\documentclass[showpacs,amssymb,twocolumn,floatfix,prl]{revtex4-1}

\usepackage{amssymb,amsfonts}
\usepackage{graphics,graphicx}

\begin{document}

\title{Strong-Coupling Theory of Counter-ions at Charged Plates}

\author{Ladislav \v{S}amaj}
\altaffiliation[On leave from ]
{Institute of Physics, Slovak Academy of Sciences, Bratislava, Slovakia}
\author{Emmanuel Trizac}
\affiliation{Universit\'e Paris-Sud, Laboratoire de Physique Th\'eorique et 
Mod\`eles Statistiques, UMR CNRS 8626, 91405 Orsay, France}

\begin{abstract}
We present an analytical approach to the 
strong coupling regime of similarly and highly charged plates in the 
presence of counter-ions. The procedure is physically transparent
and based on an exact expansion around the ground state 
formed by the two-dimensional Wigner crystal of counter-ions. 
The one plate problem is worked out, together with the two plates
situation.
Unlike previous approaches, 
the expansion is free of divergences, and is shown to be in excellent
agreement with available data of Monte-Carlo simulations under
strong Coulombic couplings. The present results shed light on the 
like-charge attraction regime.
\end{abstract}



\maketitle

The behaviour of charged particles in the vicinity of charged interfaces
is a central yet elusive problem in the equilibrium statistical 
mechanics of Coulomb fluids, including colloidal science. 
A landmark in the field was the realization in the 1980s that similarly
charged surfaces may attract each other under strong 
enough Coulombic couplings, which can be realized in practice 
increasing the valency of the counter-ions involved 
\cite{Guldbrand84,Kjellander84,Kekicheff93}. Notorious illustrations 
of this like-charge attraction are
the formation of DNA condensates \cite{Bloomfield96} or aggregates
of colloidal particles \cite{Linse00}.

The weak-coupling limit is described by the Poisson-Boltzmann mean-field 
approach \cite{Andelman06} and by its systematic improvements via 
the loop expansion \cite{Attard88,Podgornik90,Netz00}.
A remarkable achievement of the last decade has been accomplished 
in the opposite strong-coupling (SC) limit, pioneered by Rouzina and
Bloomfield \cite{Rouzina96}, substantiated by Shklovskii,
Levin and collaborators  \cite{Shklovskii,Levin02},
and formalized by Netz {\it et al} 
\cite{Moreira00,Netz01,Boroudjerdi05}.
An essential ingredient is that the layer of counter-ions close to a
charged wall becomes two-dimensional, and in the 
field-theoretical method put forward in 
\cite{Moreira00,Netz01},
the leading behaviour stems from a single-particle theory,
which produces more compact profiles than within mean-field theory
\cite{Messina09}.
Next correction orders correspond to a virial/fugacity expansion in inverse 
powers of the coupling constant $\Xi$, see the definition (\ref{3}) below. 
The method requires a renormalization of infrared divergences via 
the electroneutrality condition.
A comparison with the Monte-Carlo (MC) simulations \cite{Moreira01}
indicates the adequacy of the leading single-particle theory in the
asymptotic SC limit, while capturing the first correction 
resisted the analysis.

The establishment of an (approximative) interpolation between 
the Poisson-Boltzmann and SC regimes, based on the idea of a 
``correlation hole'',  was the subject of a series of works 
\cite{Nordholm84,Chen06,Santangelo06,Hatlo10}.
The specification of the correlation hole was done empirically in
Refs. \cite{Chen06,Santangelo06} and self-consistently, as an optimization
condition for the grand partition function, in \cite{Hatlo10}.
A relevant observation in \cite{Hatlo10}, corroborated by a comparison
with the MC simulations, was that the first correction
in the SC expansion is proportional to $1/\sqrt{\Xi}$, 
and not to $1/\Xi$ as suggested by the original SC theory. 

The aim of this Letter is to revisit the SC limit and 
establish an exact expansion 
which, in light of the previous discussion, has yet to be formulated.
The leading term of counter-ion density profiles coincides with the
single-particle picture of the SC theory.
Our expansion is free of infrared divergences and entails 
a correction in $1/\sqrt{\Xi}$ to the leading
behaviour. Our
analytical results 
are shown to be
in excellent agreement with available MC data without adjustable 
parameters. The procedure is significantly simpler than previous
works, and appears versatile. It will in particular be shown to
yield new exact results in the like charge attraction regime.

\begin{figure}
\begin{center}
\includegraphics[width=0.4\textwidth,clip]{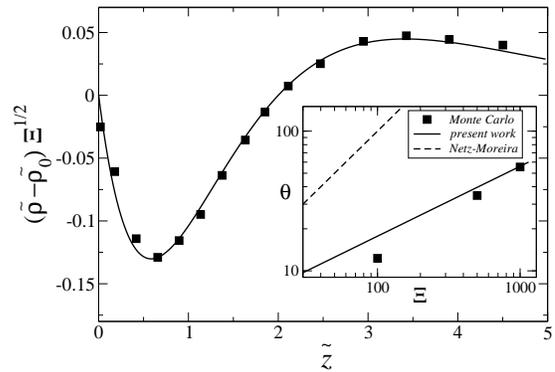}
\caption{Comparison between the analytical first correction to the
strong coupling profile (solid curve) and the Monte Carlo
results of Ref. \cite{Moreira01} at $\Xi=10^3$, 
for a single charged wall. 
The inset compares our prediction 
for the rescaling factor $\theta$ (solid curve given by 
Eq. (\ref{20})) to its Monte Carlo value reported in \cite{Moreira01}
and to the original prediction $\theta=\Xi$ (dashed line).
\label{fig:prof_theta}} 
\end{center}
\end{figure}

Here, we study a classical system of (equally charged) counter-ions in 
the vicinity of one or two planar 
walls bearing a uniform surface charge density, 
$\sigma e$ ($e$ is the elementary charge and $\sigma>0$), 
the system as a whole being electro-neutral. The system,
at thermal equilibrium at the inverse temperature 
$\beta=1/(k_{\rm B}T)$, is immersed in a 
solution of dielectric constant $\epsilon$ containing $q$-valent
counter-ions, each thus having charge $-q e$. 
For simplicity, no image forces are present.
Let us describe briefly the original SC theory \cite{Moreira00,Netz01} for 
the case of a single wall localized in the $z=0$ plane.
The counter-ions are confined to the half-space $z\ge 0$.
The relevant length scales in Gaussian units are: The Bjerrum length 
$\ell_{\rm B}=\beta e^2/\epsilon$, i.e. the distance at which two unit charges
interact with thermal energy $k_{\rm B}T$, and the Gouy-Chapman length
$\mu = 1/(2\pi q \ell_{\rm B}\sigma)$, i.e. the distance from the charged 
wall at which an isolated counter-ion has potential energy equal 
to thermal energy. 
All lengths $r$ will be expressed in units of $\mu$, $\tilde{r}=r/\mu$.
The counter-ion density profile $\rho(z)$, which only depends on the distance
from the wall $z$, will be considered in the rescaled form
\begin{equation} \label{1}
\tilde{\rho}(\tilde{z}) = \frac{\rho(\tilde{z})}{2\pi\ell_{\rm B}\sigma^2} ,
\end{equation}
so that the electro-neutrality condition 
$q \int_0^{\infty}dz \rho(z) = \sigma$
simply reads 
$\int_0^{\infty} d\tilde{z} \tilde{\rho}(\tilde{z}) = 1 .$
The coupling parameter quantifying the strength of electrostatic correlations
is 
\begin{equation} \label{3}
\Xi = 2\pi q^3 \ell_{\rm B}^2 \sigma
\end{equation}
and is large in the SC regime.
According to the SC theory \cite{Moreira00,Netz01}, the profile of the
counter-ion density can be formally expanded as
\begin{equation} \label{4}
\tilde{\rho}(\tilde{z}) = \tilde{\rho}_0(\tilde{z})
+ \frac{1}{\Xi} \tilde{\rho}_1(\tilde{z}) + {\cal O}(\Xi^{-2}) ,
\end{equation}
where
\begin{equation} \label{5}
\tilde{\rho}_0(\tilde{z}) = {\rm e}^{-\tilde{z}} , \quad
\tilde{\rho}_1(\tilde{z}) = {\rm e}^{-\tilde{z}}
\left( \frac{\tilde{z}^2}{2}-\tilde{z} \right) .
\end{equation}
The leading term $\tilde{\rho}_0(\tilde{z})$ comes from the single-particle 
picture of counter-ions in the linear surface-charge potential.
The MC simulations \cite{Moreira01} indicate that the first 
correction $\tilde{\rho}_1(\tilde{z})$ has the expected functional 
form for $\Xi>10$, however, the value of the prefactor is incorrect.
To be more particular, let us subtract the leading SC profile in (\ref{4}) 
and express the first correction as
\begin{equation} \label{6}
\tilde{\rho}_1(\tilde{z}) = \theta \left[
\tilde{\rho}(\tilde{z}) - \tilde{\rho}_0(\tilde{z}) \right] ,
\end{equation}
$\tilde{\rho}(\tilde{z})$ being the density profile obtained from 
the MC simulations. 
The factor $\theta$ can be treated as a fitting parameter which,
in the original SC theory, should be equal to $\Xi$ plus
next-to-leading corrections.
The numerical dependence of $\theta$ on $\Xi$ is pictured in 
the inset of Fig. \ref{fig:prof_theta}.
We see that $\theta$ (symbols) is much smaller than $\Xi$ (dashed line).

Our approach is based on the fact that in the asymptotic strong-coupling
limit $\Xi\to\infty$, the counter-ions collapse on the charged surface, 
creating a 2D hexagonal (equilateral triangular) Wigner crystal \cite{Levin02}
where every ion has 6 nearest neighbors forming an hexagon.
Let us denote by ${\bf R}_i=(X_i,Y_i)$ the position vectors of the vertices 
on this hexagonal lattice.
Since there are just two triangles per particle, the lattice spacing
$a$ of the globally electro-neutral structure is given by
$
q/\sigma = \sqrt{3} a^2/2.
$
Note that the strong-coupling limit coincides with the regime in which
the distance $a$ between the nearest-neighbor counter-ions is much larger
than the distance $\mu$ between the counter-ions and the charged surface
\cite{Rouzina96},
$
\tilde{a} \equiv a/\mu \propto \sqrt{\Xi} \gg 1 .
$
In the asymptotic limit $\Xi\to\infty$, each vertex ${\bf R}_i$ is occupied 
by a counter-ion $i$ $(i=1,\ldots,N; N\to\infty)$.
The ground-state energy of the counter-ion system together with the homogeneous 
background charge is $E_0$.
For $\Xi$ large but not infinite, the fluctuations of ions around their 
lattice positions start to play a role.

Let us first shift one of the particles, say $i=1$, from its lattice 
position ${\bf R}_1$ by a small vector $\delta{\bf R}_1=(x,y,z)$ 
($\delta R_1 \equiv \vert \delta{\bf R}_1 \vert \ll a$) and look for 
the corresponding change in the total energy $\delta E = E - E_0\ge 0$.
The first contribution to $\delta E$ comes from the interaction of the 
shifted counter-ion with the potential induced by the homogeneous surface 
charge density:
$\delta E^{(1)} = 2\pi q e^2 \sigma z /\epsilon.
$
The second contribution to $\delta E$ comes from the interaction of 
the particle with all other particles on the 2D hexagonal lattice.
This contribution can be expanded as an infinite series in $x$, $y$ and $z$;
for our purposes, it is sufficient to consider this expansion up to
harmonic terms, which, in the $z$-direction, read
\begin{equation} \label{10}
\epsilon\delta E^{(2)}_z = \sum_{i\ne 1} \left[ \frac{(q e)^2}{\sqrt{R_i^2+z^2}}
- \frac{(q e)^2}{R_i} \right] \sim - \frac{(q e)^2}{2 a^3} S z^2. 
\end{equation}
Here, the dimensionless quantity 
$S = \sum_{i\ne 1} (R_i/a)^{-3}$ can be expressed 
from the general theory of lattice sums \cite{Zucker74,Zucker75}
\begin{eqnarray}
S & = & \sum_{m,n=-\infty\atop (m,n)\ne (0,0)}^{\infty}
\frac{1}{(m^2+m n+n^2)^{3/2}} \nonumber \\
& = & \frac{2}{\sqrt{3}} \zeta\left( \frac{3}{2}\right) \left[
\zeta\left( \frac{3}{2},\frac{1}{3}\right) -
\zeta\left( \frac{3}{2},\frac{2}{3}\right) \right] , \label{11}
\end{eqnarray} 
where $\zeta(z,q)=\sum_{n=0}^{\infty} 1/(q+n)^z$ is the generalized Riemann 
zeta function \cite{Gradshteyn} and $\zeta(z)\equiv\zeta(z,1)$.
Explicitly, $S=11.034\ldots$.
A shift of the particle simultaneously along all directions does not
induce ``mixed'' harmonic terms of type $xz$ or $yz$.
The harmonic term in the $(x,y)$-plane can be computed, and
in dimensionless form, we have
\begin{equation} \label{14}
-\beta \delta E \sim - \tilde{z} + \frac{3^{3/4}}{(4\pi)^{3/2}}
\frac{S}{\sqrt{\Xi}} \left[ \frac{\tilde{z}^2}{2} - \frac{1}{4}
\left( \tilde{x}^2+\tilde{y}^2 \right) \right] .
\end{equation}
This formula reveals a relationship between the order of the expansion
of $-\beta \delta E$ in the dimensionless lengths 
$\tilde{x},\tilde{y},\tilde{z}$ and the SC expansion in $1/\sqrt{\Xi}$.
The linear term $-\tilde{z}$, which is the only one which does not vanish
in the limit $\Xi\to\infty$, is the leading term.
It corresponds to the single-particle picture, in close analogy
with the previous SC theory.
The harmonic terms turn out to be of the SC order
$\beta (q e)^2 \mu^2 /a^3\propto 1/\sqrt{\Xi}
$
and likewise, terms of the $p$th order in the variables
$\tilde{x},\tilde{y},\tilde{z}$ are of the SC order
$
\beta (q e)^2  \mu^p / a^{p+1} \propto 1/\Xi^{(p-1)/2} .
$
This scheme constitutes a systematic basis for SC expansions.

The generalization of the above formalism to all particles is straightforward.
We shift every particle $i=1,2,\ldots,N$ from its lattice position 
${\bf R}_i$ by a small vector $\delta{\bf R}_i=(x_i,y_i,z_i)$. 
In what follows however, 
we shall be interested in the counter-ion density profile
which only depends on the $\tilde{z}$ coordinate.
Thus, when expanding in statistical averages the Gibbs weight
$\exp(-\beta\delta E)$ in powers of $1/\sqrt{\Xi}$, we can restrict 
ourselves to the $z$-harmonic part.
The corresponding change in the total energy $\delta E$ is given by 
a counterpart of (\ref{14}),
\begin{equation}
-\beta \delta E  \sim  - \sum_i \tilde{z}_i + 
\frac{3^{3/4}}{16 \pi^{3/2}} \frac{1}{\sqrt{\Xi}} 
\sum_{i<j} \frac{(\tilde{z}_i-\tilde{z}_j)^2}{
(\vert {\bf R}_i-{\bf R}_j \vert/a)^3} .
\label{17}
\end{equation}
The next simplification comes from the fact that particles are identical,
exposed to the same potential induced by the surface charge, so that
a summation over particle coordinates can be represented by just one 
auxiliary coordinate.
For the density particle profile, defined by 
$\rho({\bf r}) = \langle \sum_{i=1}^N \delta({\bf r}-{\bf r}_i) \rangle =
N \langle \delta({\bf r}-{\bf r}_1) \rangle$, we get explicitly
\begin{equation}
\tilde{\rho}(\tilde{z})  =   C {\rm e}^{-\tilde{z}} \int_0^{\infty} d\tilde{z}'
{\rm e}^{-\tilde{z}'} 
\Bigg[ 1 + \frac{3^{3/4}S}{16\,\pi^{3/2}} 
\frac{(\tilde{z} - \tilde{z}')^2}{\sqrt{\Xi}}
\Bigg] + {\cal O}(\frac{1}{\Xi}) , \label{18}
\end{equation}
where $C$ is determined by the normalization condition $\int \tilde{\rho}=1$.
Simple algebra gives
\begin{eqnarray}
\tilde{\rho}(\tilde{z}) & = &  {\rm e}^{-\tilde{z}} + 
\frac{3^{3/4}}{(4\pi)^{3/2}} \frac{S}{\sqrt{\Xi}} {\rm e}^{-\tilde{z}} 
\left( \frac{\tilde{z}^2}{2} - \tilde{z} \right) + {\cal O}(\Xi^{-1}) . 
\nonumber \\ & & \label{19}
\end{eqnarray}
Comparing this result with the previous one (\ref{4}), (\ref{5}) obtained
in the original SC theory \cite{Moreira00,Netz01}, we see that the leading 
terms coincide, while the first corrections have the same functional 
dependence in space but different prefactors.
The result (\ref{19}) can be re-expressed in terms of the $\theta$-factor,
introduced in the relation (\ref{6}), as follows
\begin{equation} \label{20}
\theta = \frac{(4\pi)^{3/2}}{3^{3/4}} \frac{1}{S} \sqrt{\Xi}
= 1.771\ldots \sqrt{\Xi} .
\end{equation}
This formula, in excellent agreement with
MC data, differs substantially from the previous SC estimate 
$\theta=\Xi$, see Fig. \ref{fig:prof_theta}.

\begin{figure}[htb]
\begin{center}
\includegraphics[width=0.45\textwidth,clip]{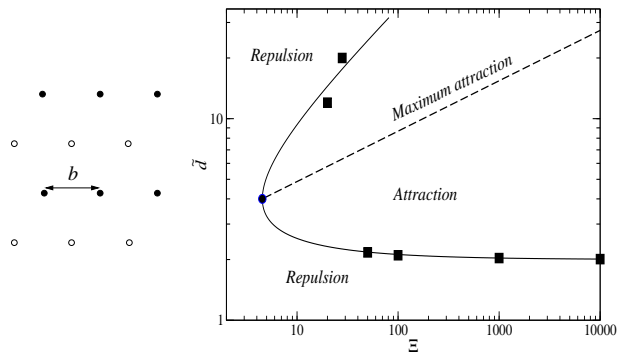}
\caption{Left: Structure I of counter-ions on two parallel charged plates
(see text).
Right: Phase diagram following from the equation of state (\ref{eq:eos}):
the solid curve shows the points where $P=0$.
The filled squares are those MC results from Ref \cite{Moreira01}
with $\Xi>20$, while 
the dashed line is for the points where 
$\partial\tilde{P}/\partial\tilde{d}=0$, which defines 
$\tilde d_{\text max}$.
\label{fig:phases}}
\end{center}
\end{figure}

The method can be readily applied to the case of two parallel walls, each having 
the same charge density $\sigma e$, located at distance $d$ from one another. 
The electric field between the walls is equal to 0 now. 
At $T=0$, the classical system is defined by the dimensionless separation 
$\eta=d\sqrt{\sigma/q}=\tilde{d}/\sqrt{2\pi\Xi}$. 
A complication comes from the fact that counter-ions form, on the opposite 
surfaces, a bilayer Wigner crystal, the structure of which depends on $\eta$ 
\cite{Goldoni96,Messina03,Lobaskin07}.
Five different structures are energetically favored for various
regions of $\eta$.
At the smallest separations  when $\eta \to 0$, 
the single hexagonal lattice (structure I, see Fig. \ref{fig:phases}-left), 
with rows 
distributed consecutively between the two surfaces, is formed.
At large separations $\eta>0.732$, the energetically favored geometry
is composed of two staggered hexagonal lattices (structure V), one for each plate. 
We shall document our SC approach on structure I,
relevant to obtain the large $\Xi$ behaviour.
Due to global neutrality, the lattice spacing $b$ of the single (bilayer) 
hexagonal structure is given by
$q/(2\sigma) = \sqrt{3}b^2 /2.$
The SC regime is identified with the condition $d\ll b$, or equivalently
$\Xi \gg \tilde{d}^2$.

The two walls are located at positions $z=0$ and $z=d$.
The position vector ${\bf R}_i$ of the particle localized on the shared 
hexagonal Wigner lattice will be denoted as ${\bf R}^{(0)}_i$ if it belongs 
to the wall at $z=0$ (say filled symbols of the left panel of
Fig. \ref{fig:phases}) and as ${\bf R}^{(d)}_i$ 
if it belongs to the wall at $z=d$ (open symbols in Fig. \ref{fig:phases}).
Let us shift the particle $i=1$ localized on the $z=0$ wall by a small
vector $\delta {\bf R}_1^{(0)}=(x,y,z)$ and look for the energy change
$\delta E$ from the ground state.
Since the potential induced by the surface charge on the walls is
constant between the walls, the corresponding $\delta E^{(1)}=0$.
The harmonic term in the $z$-direction reads
\begin{equation} \label{22}
\epsilon\delta E_z^{(2)} = \frac{(q e)^2}{2 b^3} \left[
- \sum_{i\ne 1} \frac{z^2}{(R_i^{(0)}/b)^3} +
\sum_{i} \frac{d^2-(d-z)^2}{(R_i^{(d)}/b)^3} \right] .
\end{equation}
Using the exact values of the partial hexagonal sums \cite{Zucker75} 
$\sum_{i\ne 1} \left[b/R_i^{(0)}\right]^3 = 5S/12 $,
$\sum_{i} \left[b/R_i^{(d)}\right]^3 = 7S/12 $,
$\delta E_z^{(2)}$ turns out to be positive, as it should.
The harmonic term in the $(x,y)$-plane can again be computed
but proves immaterial for the sake of our purposes.
When all particles are shifted from their lattice positions 
$\{ {\bf R}_i \}$ by $\{ \delta {\bf R}_i = (x_i,y_i,z_i) \}$,
the total energy change is given, as far as the $z$-dependent contribution 
is concerned, by
\begin{eqnarray}
-\beta \delta E & \sim & 
- \frac{3^{3/4}}{(4\pi)^{3/2}} \frac{\sqrt{2}}{\sqrt{\Xi}} \frac{1}{2}
\sum_{i,j} \frac{\tilde{d}^2}{ 
(\vert {\bf R}_i^{(0)}-{\bf R}_j^{(d)} \vert/b)^3} \nonumber \\ & & 
+ \frac{3^{3/4}}{(4\pi)^{3/2}} \frac{\sqrt{2}}{\sqrt{\Xi}}  \frac{1}{2} \sum_{i<j} \frac{(\tilde{z}_i-\tilde{z}_j)^2}{
(\vert {\bf R}_i-{\bf R}_j \vert/b)^3}  . 
\label{24}
\end{eqnarray}
Expanding $\exp(-\beta\delta E)$ in $1/\sqrt{\Xi}$
and enforcing electro-neutrality, the density profile 
$\tilde{\rho}(\tilde{z})$ is obtained in the form
\begin{eqnarray} 
\label{25}
&&\tilde{\rho}(\tilde{z}) = \frac{2}{\tilde{d}} + \frac{1}{\theta}
\frac{2}{\tilde{d}} \left[ \left( \tilde{z}-\frac{\tilde{d}}{2} \right)^2
- \frac{\tilde{d}^2}{12} \right] + {\cal O}(\Xi^{-1}) ~
\\
&& \hbox{where }\quad
\label{26}
\theta = \frac{(4\pi)^{3/2}}{3^{3/4}} \frac{1}{S} \frac{\sqrt{\Xi}}{\sqrt{2}}
= 1.252\ldots \sqrt{\Xi} .
\end{eqnarray}
This $\theta$ differs from the single-plate one (\ref{20}) by 
the factor $1/\sqrt{2}$ due to the different hexagonal lattice
spacings $a$ and $b$.
The functional form of (\ref{25}) coincides with that of Moreira and Netz
\cite{Moreira00,Netz01}.
For (not yet asymptotic) $\Xi=100$, the previous SC result $\theta=\Xi$ 
is far away from the MC estimate $\theta\simeq 11.2$
\cite{Moreira01}, while our formula (\ref{26}) gives 
$\theta\simeq 12.5$.

Applying the contact-value theorem to the density profile (\ref{25}), 
the pressure $P$ between the plates is given by 
\begin{equation} \label{eq:eos}
\tilde{P} \,=\, \frac{P}{2 \pi \ell_B \sigma^2} \,=\,
- 1 + \frac{2}{\tilde{d}} + \frac{\tilde{d}}{3\theta}
+ {\cal O}\left(\frac{\tilde{d}^2}{\Xi}\right) .
\end{equation}
An analogous result was obtained within the approximate approach 
of Ref. \cite{Hatlo10}, with the underestimated ratio 
$\theta/\sqrt{\Xi}=3\sqrt{3}/2=0.866\ldots$.
We recall that our estimate of $\theta$ is valid only in the structure
I region $0\le \eta<0.006$; increasing $\eta$, other energetically favored 
bilayer Wigner structures have to be considered as a starting point.
Eq. (\ref{eq:eos}) provides insight into the like charge 
attraction phenomenon.
The attractive ($P<0$) and repulsive
($P>0$) regimes are shown in Fig. \ref{fig:phases} (right panel).
Although our results
hold for $\tilde{d} \ll \Xi^{1/2}$ and for large $\Xi$, the shape
of the phases boundaries where $P=0$ (solid curve) shows striking
similarity with its counterpart obtained numerically 
\cite{Moreira01,Chen06} (we note for instance that the 
terminal point shown by the filled circle
in Fig \ref{fig:phases} is located at $\tilde d=4$, a value close
to that which can be extracted from \cite{Moreira01,Chen06}). While the upper
branch of the attraction/repulsion boundary is such that 
$\tilde d/\sqrt{\Xi}$ is of order unity
and hence lies at the limit of validity of our expansion, we predict the
maximum attraction to be obtained for $\tilde d_{\text{max}} = \sqrt{6 \theta}
\propto \Xi^{1/4}$, as follows from enforcing 
$\partial\tilde{P}/\partial\tilde{d}=0$. Since 
$\tilde d_{\text{max}}/\sqrt{\Xi} \propto \Xi^{-1/4} \to 0$, we can consider
the latter prediction, shown by the dashed line in Fig. \ref{fig:phases},
as asymptotically exact; we note that it is fully corroborated by the 
scaling laws reported in \cite{Chen06}.

In conclusion, the present exact approach to the strong coupling regime 
shows that while the leading order results at large $\Xi$ (for one or two 
plates)
can be obtained by a single counter-ion theory, the next terms  actually
reflect the complete ground state structure ($N$ counter-ion property).
This explains the failure of virial-like expansions. 
We have shown how such shortcomings can be circumvented
within a physically transparent procedure, and 
obtained analytical results in remarkable agreement with Monte Carlo data.
In practice, for a highly charged interface in water at room $T$,
one has $\sigma \ell_{\rm{B}}^2 \simeq 1$, so that $\Xi$ takes values close
to 50, 170 and 400 for respectively 
di-, tri-, and tetra-valent counter-ions, as often found in biology
(spermine). Although asymptotic, our predictions turn out to be
reliable for such couplings.
A generalization of the approach to dielectric inhomogeneities
\cite{Jho08}, systems with salt or asymmetric \cite{Kanduc10},
and curved surfaces \cite{Santos09}, offer interesting 
problems for more detailed studies in the future.
The formulation is also convenient for a quantum-mechanical 
generalization.

\begin{acknowledgments}
The support received from Grant VEGA No. 2/0113/2009 and CE-SAS QUTE 
 is acknowledged. 
\end{acknowledgments}

\end{document}